# Mid-infrared monocrystalline interference coatings with excess optical loss below 10 ppm

G. WINKLER,[1*] L. W. PERNER,[1*] G.-W. TRUONG,[2,3] G. ZHAO,[4] D. BACHMANN,[2] A. S. MAYER,[1] J. FELLINGER,[1] D. FOLLMAN,[2,3] P. HEU,[2] C. DEUTSCH,[2] D. M. BAILEY,[4] H. PEELAERS,[5] S. PUCHEGGER,[6] A. J. FLEISHER,[4] G. D. COLE,[2,3] O. H. HECKL[1]

[1]*Christian Doppler Laboratory for Mid-IR Spectroscopy and Semiconductor Optics, Faculty Center for Nano Structure Research, Faculty of Physics, University of Vienna, Boltzmanngasse 5, 1090 Vienna, Austria*
[2]*Crystalline Mirror Solutions, Santa Barbara, CA and Vienna, Austria*
[3]*Thorlabs Crystalline Solutions, 114 E Haley St., Suite G, Santa Barbara, CA 93101 USA*
[4]*Material Measurement Laboratory, National Institute of Standards and Technology, 100 Bureau Drive, Gaithersburg, MD 20899 USA*
[5]*Department of Physics & Astronomy, University of Kansas, 1251 Wescoe Hall Dr., Lawrence, KS 66045 USA*
[6]*Faculty Center for Nano Structure Research, Faculty of Physics, University of Vienna, Boltzmanngasse 5, 1090 Vienna, Austria*

*\*These authors contributed equally to this work. Corresponding author: georg.winkler@univie.ac.at*

**We present high-reflectivity substrate-transferred single-crystal GaAs/AlGaAs interference coatings at a center wavelength of 4.54 μm with record-low excess optical loss below 10 parts per million. These high-performance mirrors are realized via a novel microfabrication process that differs significantly from the production of amorphous multilayers generated via physical vapor deposition processes. This new process enables reduced scatter loss due to the low surface and interfacial roughness, while low background doping in epitaxial growth ensures strongly reduced absorption. We report on a suite of optical measurements, including cavity ring-down, transmittance spectroscopy, and direct absorption tests to reveal the optical losses for a set of prototype mirrors. In the course of these measurements, we observe a unique polarization-orientation-dependent loss mechanism which we attribute to elastic anisotropy of these strained epitaxial multilayers. A future increase in layer count and a corresponding reduction of transmittance will enable optical resonators with a finesse in excess of 100 000 in the mid-infrared spectral region, allowing for advances in high resolution spectroscopy, narrow-linewidth laser stabilization, and ultrasensitive measurements of various light-matter interactions.**

## 1. INTRODUCTION

High-performance mirrors are employed for the construction of optical resonators in a variety of applications in optics and photonics. Stable resonators are routinely used to narrow the linewidth of continuous-wave lasers, thereby creating optical references for frequency comb stabilization and precision molecular spectroscopy [1–3]. Stable interferometers can be very small or very large in size, enabling scientific discovery in fields as seemingly unrelated as microcavity sensing [4] and gravitational wave detection [5]. Emerging applications in chemical sensing [6], discrete imaging [7], ultracold chemistry [8,9], and even fundamental physics [10] would benefit immediately from high-performance mirrors at mid-infrared (mid-IR) wavelengths (loosely defined here as the spectral range from 3 μm to 12 μm) to probe new and interesting phenomena with increased precision. A long-standing goal is the development of low-loss mirrors such as those readily available throughout the near-infrared (near-IR) spectral region.

Traditional physical vapor deposition (PVD) techniques for the fabrication of high-reflectivity "supermirrors" such as ion-assisted evaporation or ion-beam sputtering are not widely available in the mid-IR or have not been optimized for the use of materials transparent in this spectral range. The predominant process used for such optics is traditional electron beam or thermal evaporation. Thus, the excess optical loss, that is the combined optical power absorption and scattering losses ($L = A + S$), of those mid-IR mirrors are generally > 100 parts-per-million (ppm) [11], with only rare exceptions [12]. High-quality near-IR mirrors on the other hand are routinely capable of excess optical loss at the < 10 ppm level.

Excess loss strongly influences the on-resonance transmission $T_{\text{cav}}$ of a cavity, and thereby the signal-to-noise ratio (SNR) in cavity-enhanced spectroscopy efforts, according to

$$T_{\text{cav}} = \frac{T^2}{(T+L)^2}. \qquad (1)$$

Here we assume equal power transmission coefficients $T$ for both mirrors of a linear cavity and perfect mode matching. Simultaneously, one typically tries to maximize the cavity finesse

$$\mathcal{F} = \frac{\pi}{T + L}, \quad (2)$$

which is inversely proportional to the cavity total loss $T + L$. Total loss can also be written as $1-R$, with $R$ denoting the power reflection coefficient of a mirror, as a direct consequence of conservation of energy $R + T + A + S = 1$. While $T$ is a design parameter that is controllable via the layer count of an interference coating, $L$ is usually limited by the choice of materials and fabrication technology. If $L$ is non-negligible in comparison to $T$, it is therefore necessary to strike a compromise in the maximization of both the above figures of merit. Sacrificing $T$ for lower total losses then often results in cavities of both moderate $\mathcal{F}$ around 25 000 – 30 000 and $T_{cav}$ well below 10%.

Substrate-transferred monocrystalline interference coatings are a promising solution to these challenges in the mid-IR spectral region. With a room-temperature transparency window that extends from approximately 0.87 μm to beyond 10 μm, alternating multilayers of high refractive index gallium arsenide (GaAs) and low refractive index ternary aluminum gallium arsenide ($Al_xGa_{1-x}As$) alloys can, in principle, be used to create high-reflectivity coatings over a broad wavelength window without fundamental adjustments in the manufacturing process. Crystalline coatings were originally developed as a means to overcome the Brownian-noise limit in precision interferometry [13], having simultaneously achieved ppm levels of optical losses and elastic loss angles in the $10^{-5}$ range [14].

Rather than relying on direct deposition onto the final optical substrate, these coatings are generated using an epitaxial layer transfer process. Initially, a monocrystalline heterostructure is grown on a lattice-matched GaAs base wafer via molecular beam epitaxy (MBE). Following the crystal growth process, a microfabrication procedure is employed in order to transfer the epitaxial multilayer to arbitrary optical substrates, including curved surfaces, via direct bonding. This technique yields monocrystalline interference coatings with high purity, low defect density, abrupt interfaces, and low surface roughness, which in turn enables high-reflectivity (HR) coatings with state-of-the-art optical absorption of $A < 1$ ppm from 1.0 μm to 1.6 μm, with optical scatter of $S < 3$ ppm in the same near-IR wavelength range [15].

In practice, however, extending the performance of monocrystalline coatings into the mid- and far-infrared is not trivial. The first prototype mid-IR mirrors at design wavelengths of 3.3 μm and 3.7 μm exhibited promising excess loss levels of 159 ppm and 242 ppm, respectively [15]. These results were on par with other state-of-the-art mid-IR coatings, but still not comparable to their state-of-the-art near-IR counterparts capable of $A + S < 10$ ppm. Regardless, owing to the relatively low losses, these prototype mirrors helped enable the first detection and characterization of the transient intermediate radical *trans*-DOCO in the atmospheric-pressure reaction of the deuterated hydroxyl radical, OD, with carbon monoxide, CO [16]. Despite their vital role in studying real-time atmospheric chemical kinetics, however, the excess optical losses of these first prototypes were limited by excess scatter driven by structural defects in the very thick (20 – 30 μm) epitaxial multilayers.

Following significant improvements in the mid-IR crystalline coating production process as outlined in this work, we report on monocrystalline interference coatings with $A + S \leq 10$ ppm at a wavelength of 4.54 μm, thus demonstrating ultralow-loss mirrors with breakthrough performance in the mid-IR. These mirrors now realize the full performance potential predicted for the level of crystal purity feasible in a state-of-the-art MBE process (see Fig. 7 in Ref. [15]), at the longest wavelength demonstrated to date.

In the course of this work we independently characterize the wavelength- and polarization-dependent mirror metrics using a comprehensive suite of advanced, home-built optical metrology tools. Several of those setups have been explicitly designed to work with a single broadband Fabry–Pérot quantum cascade laser (FP-QCL) as an optical probe, being low-cost and available over a very wide range of mid-IR wavelengths, such that the characterization can be easily extended to future mirrors with different design wavelengths (a similar approach for ringdown measurements in the near-IR was presented in [17]).

The demonstration of mid-IR mirrors with a fifteen-fold reduction in $A + S$ will allow for improved sensitivity in a variety of linear and nonlinear cavity-enhanced spectroscopy techniques. Specifically, the ultra-low-loss mirrors reported here were designed for a target wavelength of 4 500 nm to explore next-generation benchtop optical instruments for the detection of radiocarbon dioxide ($^{14}CO_2$) and rare clumped isotopic substitutions of nitrous oxide ($N_2O$). To date, optical detection of $^{14}CO_2$ by saturated absorption cavity ring-down spectroscopy (SCAR) has achieved sensitivity levels comparable with accelerator mass spectrometry facilities [18], and several competing linear absorption sensor architectures have also been demonstrated with known positive and negative trade-offs related to sensitivity [19,20], accuracy [21], and portability [22,23]. Recently, a scheme for Doppler-free two-photon cavity ring-down spectroscopy (TP-CRDS) of $N_2O$ was demonstrated at 4.53 μm with a projected detection limit that was substantially better than currently available commercial gas analyzers [24] and its application to the sensitive and selective detection of $^{14}CO_2$ has also been proposed [25]. These immediate applications in trace gas and rare isotope detection using high-finesse optical resonators, as well as the previously mentioned potential applications in time-resolved spectroscopy, ultracold chemistry, and fundamental physics, make the advanced fabrication of ultralow-loss mid-IR mirrors of broad interest to the optics and photonics community.

## 2. EXPERIMENTAL DETAILS

As this was a first attempt at fabricating crystalline HR coatings with a center wavelength longer than 4.0 μm, as well as using an improved production process for minimizing excess optical losses, we undertook a comprehensive investigation of the mirror performance. In the course of our characterization efforts we have determined the mirror reflectance, R, via cavity ring-down; the transmittance, T, by directly probing the transmission through the mirror (complemented by detailed calculations); and the polarization-dependent relative absorption, ΔA, employing photothermal common-path interferometry (PCI) [26]. This enabled us to extract each individual loss component, as the magnitude of A can be inferred via the aforementioned expression R + T + A + S = 1. For these mirrors, scatter, S, is assumed to be negligible based on the observation of S + A < 3 ppm of similar crystalline mirrors in the near-IR [15] and decreasing scatter loss with increasing wavelength. For example, a surface roughness below 0.2 nm, as is routinely achieved with these optimized coatings, results in a calculated scatter loss of ~ 0.3 ppm at 4540 nm [27].

## Mirror design and transmission model refinement

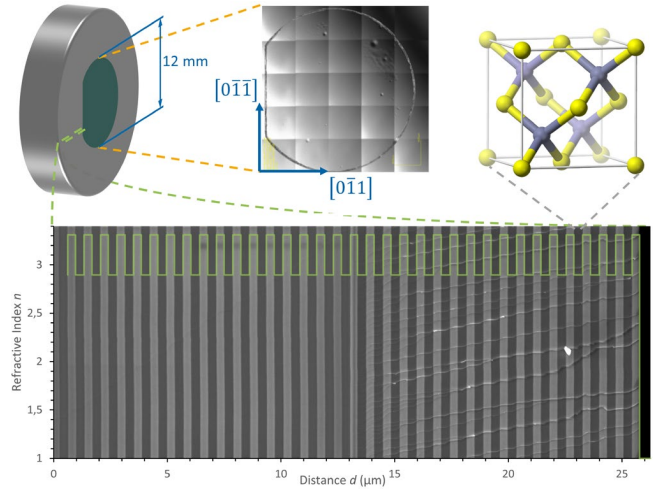

**Fig. 1.** Details of the prototype mirror structure. Top left: Schematic of the 12 mm diameter crystalline coating bonded onto a Si substrate of 25.4 mm diameter with a concave 1 m radius of curvature. The flat is aligned with the $[0\bar{1}1]$ crystal axis. Top middle: Stitched optical micrographs of the 12 mm diameter coating surface following the substrate transfer process. Top right: Zincblende structure of the crystalline coating (Ga in yellow, As in gray). Bottom: Cross-sectional SEM images of the layer stack consisting of 34.5 periods of alternating GaAs (light gray) and AlGaAs (dark gray) layers. The full coating was fabricated by stacking two intermediate structures grown with 17 periods capped with a 1/8th wave GaAs layer prior to substrate transfer. The bonding line is faintly visible in the middle of the centermost GaAs layer. Abrupt interfaces allow for precise measurements of individual layer thicknesses via edge detection (The green line indicates the refractive index profile). The terraced cracks in the cleaved coating (visible in the right half of the multilayer) result from a minute mismatch in crystal axis orientation from the stacking process, leading to an imperfect cleave of the top half of the mirror.

The reflectance and transmittance of multilayer thin film structures can readily be calculated via transfer matrix methods (TMM) [28]. The quarter-wave layer structure of the crystalline mid-IR mirrors under study was designed for a stopband center wavelength of 4 500 nm with a target transmittance of 140 ppm, comprising 34.5 periods of GaAs/Al$_{0.92}$Ga$_{0.08}$As with individual layer thicknesses of 340.0 nm and 388.6 nm respectively, based on refractive index values taken from Afromowitz et al. [29]. Deposition of the crystalline interference coating was carried out in a multi-wafer MBE system employing a 14×4" wafer configuration with on-axis (100)-oriented 100 mm diameter semi-insulating GaAs base wafers. For these mirrors, we employed a novel fabrication process to minimize the limiting optical scatter. This was realized by reducing the total thickness of the grown heterostructure by using stacked optical coatings [30]. In this process we bond two "half mirrors", halving the epitaxial multilayer thickness and significantly improving the material surface quality. In this process the potentially defective free surface of the as-grown crystal is embedded in the middle of the multilayer and the face of the mirror that directly interacts with the optical field exhibits a substantially improved surface quality, as with our flipped optical coatings described in Ref. [15]. Following the MBE growth process, we ran a wafer-scale GaAs-to-GaAs bonding process, followed by substrate and etch stop removal on one of the wafer pairs, to generate the full coating stack. The 12 mm diameter coating discs were then lithographically patterned and selectively etched through the stacked epitaxial multilayer in preparation for the substrate transfer process. Finally, the coating discs were transferred via a second direct bonding process to 25.4 mm diameter silicon (Si) substrates with a concave 1 m radius of curvature and 6.35 mm thickness. The planar backside of each substrate was coated with a broadband (standard PVD) anti-reflection (AR) coating over the range of 3 μm to 5 μm ($R$ = 0.3% at 4.5 μm). Both standalone coating discs and completed mirrors on Si substrates were used for the characterization efforts described in the course of this manuscript.

To begin the analysis process, variations in the deposition rate during the MBE growth process necessitate correction of these nominal layer thicknesses to more accurately represent the as-grown samples. These corrections were determined using a combination of X-ray diffraction (XRD), Fourier transform spectrometry (FTS, Bruker VERTEX 80), and cross-sectional scanning electron microscopy (SEM, Zeiss Supra 55VP). In addition to providing guidance on the layer thicknesses, by probing the material lattice constant, XRD measurements also provide an estimate of the alloy composition of the low index ternary AlGaAs alloy with relative error bounds on the order of 1%.

Cross-sectional SEM imaging of a cleaved interference coating stack prior to the substrate transfer process allowed us to determine the as-grown layer thicknesses (see Fig. 1) by means of digital post-processing using an edge-detection and peak-finding routine (ImageJ, IJ BAR package). It is expected that the derived layer thicknesses differ among mirrors of the same production batch by a global scaling factor, depending on the exact lateral position inside the MBE chamber, while the relative layer thicknesses are expected to be very similar over a wide area [31]. The relative error in SEM length measurement calibration was estimated to be 1%, determined by measuring a calibration sample (Raith CHESSY). Additionally, edge detection introduced an error of ± 4 nm due to the limited resolution of the SEM images.

Broadband FTS measurements generated transmission spectra of the mirrors under test with spectral resolution of ± 0.5 nm and were recorded in a nitrogen atmosphere at ambient pressure. While the FTS does not allow for direct measurement of the mirror transmittance close to the mirror center wavelength due to its limited sensitivity and low SNR, the stopband width and characteristic structure of the side-lobes in these spectra enable a precise extrapolation to the center wavelength through a model fit. The TMM model was fitted to the FTS spectra using the SEM-determined layer thicknesses as starting values, with the aforementioned global scaling factor, and the Al concentration in the low index layers (bounded by XRD error intervals) as free parameters.

To obtain an error estimate of the derived mirror transmittance at the center wavelength, the following Monte-Carlo-style procedure was employed: Starting from the above initial fit values, we performed further forward model calculations to sample the space of reasonable transmission spectra. Candidate spectra were generated for random variations of the best-fit parameters (within their respective relative error bounds). From these candidates, we expunged all those where center wavelength and FWHM of the mirror stopband deviated by more than ± 0.5 nm from the best initial fit, thereby excluding input parameter combinations that lead to an unphysical deviation from FTS measurements. The set of remaining spectra was then used to determine mean and standard error (at the ppm level) of the minimum transmittance. Note that in this procedure we have assumed abrupt interfaces and no variation in Al content over the full structure, as well as a perfect AR coating on the backside of the Si substrate. The refractive indices for the epitaxial materials were taken from [29], while for the Si substrate the dispersion data is taken from [32]. A refractive index of $n$ = 1 was assumed for the incident and exit media.

**Cavity Ring-down Measurements**

We constructed a linear resonator from two mirrors of the same production batch and implemented two variations of the well-established cavity ring-down technique [33] to infer the total loss, $1-R$, independent of laser source amplitude fluctuations, by measuring the cavity decay time constant, $\tau$, and cavity length, $d$, and using the equation $(1-R) = d/(c\tau)$, where $c$ is the speed of light.

At the Christian Doppler Laboratory for Mid-IR Spectroscopy and Semiconductor Optics (CDL) in Vienna, Austria, a low-cost broadband FP-QCL (TL QF4550CM1) was coupled into multiple longitudinal modes of the cavity by exploiting direct passive feedback in a simple linear configuration (Fig. 2). The sample mirrors were mounted $30.6 \pm 0.2$ cm apart as end elements of a custom vacuum chamber, which was first purged with nitrogen, then evacuated to typical pressures of 0.5 kPa to 1.0 kPa using an oil-free roughing pump. Ring-down time-traces were recorded using an amplified InAsSb detector (PDA10PT-EC). A custom microcontroller-based threshold detection circuit was used to detect on-resonance cavity transmission, shutter the laser, and trigger data capture [34]. Spectral coverage extends from about 4 520 nm to 4 720 nm (2 119 cm$^{-1}$ to 2 212 cm$^{-1}$), with a monochromator (Spectral Products Digiköm CM 110, grating 300G/mm, blaze wavelength 2.5 µm) enabling measurements at 5 nm spectral resolution. The peak of the overall FP-QCL emission bandwidth was steered via additional external grating feedback (first order reflected backwards) and monitored by a custom low-resolution FTS instrument. With the FP-QCL and sample mirrors forming a coupled cavity system, a delay stage between the source and backside of the first mirror was tuned to find resonance conditions causing power buildup in the cavity formed by the sample mirrors to occur within the spectral transmission window of the monochromator. During measurements, the transmitted TEM$_{00}$ transverse cavity modes were monitored on a mid-IR microbolometer camera (Visimid Phoenix) to restrict sampling to the smallest possible area on the mirrors and to avoid transversal mode beating. Given the cavity length resulting in a free spectral range of $\nu_{FSR} = 490 \pm 5$ MHz, beat notes between multiple longitudinal modes could not be resolved in the measurements.

To avoid birefringence-induced polarization mode beating, the linear polarization of the FP-QCL was aligned parallel to the net slow or fast axis of the ring-down cavity and exponential fits to decay curves checked for clean modulation-free residuals. Since the monochromator grating acted as a fixed polarization analyzer, it was not possible to study polarization-dependent effects at arbitrary input polarization angles.

We analyzed the theoretical possibility of a systematic bias of the measured ringdown times through correlations between the intracavity power spectrum and etalon resonances. We experimentally verified that influences of a potential etalon between FP-QCL end facet and ring-down cavity are in fact averaged out through the broadband excitation, by introducing varying levels of additional losses (up to 50%) in between these two elements, without observing systematic effects.

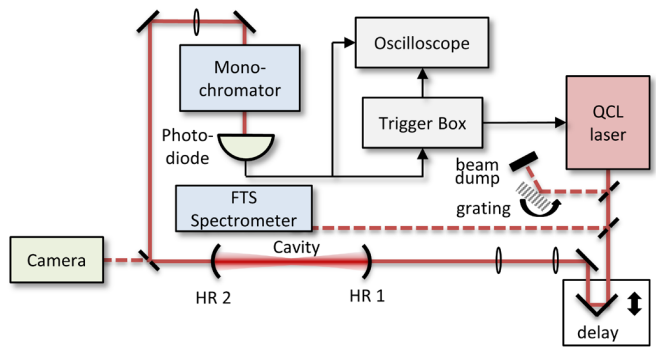

**Fig. 2.** Schematic of the FP-QCL based ring-down setup at CDL. In this setup a FP-QCL without internal stabilization is utilized for excitation, with a custom trigger circuit initiating the ring-down process. This architecture allows for significant flexibility in mirror characterization via changes in the laser source. This system is similar to that described in [17] with additional wavelength monitoring, while operating in the mid-IR and without raster-scanning capabilities. To control the center of the emission spectrum, additional feedback to the FP-QCL is provided via a reflection grating in Littrow configuration.

An independent experiment at the National Institute of Standards and Technology (NIST) in Gaithersburg, Maryland, USA was performed to cross-check the FP-QCL-based broadband measurements at CDL and to evaluate potential systematic errors in the experimental determination of the quantity $1-R$. There, a continuous-wave external-cavity QCL (EC-QCL) with linewidth < 10 MHz, capable of exciting individual longitudinal modes of a linear cavity, was used to measure $1-R$ as a function of wavelength from 4 480 nm to 4 600 nm (2 174 cm$^{-1}$ to 2 232 cm$^{-1}$) with high resolution, as illustrated in Fig. 3. At NIST, the EC-QCL was passively coupled to a single mode of the cavity by dithering the laser current, an operation which also acted as an optical shutter to initiate decay events. Finally, using either a quarter-wave or half-wave plate and a linear polarization analyzer, the EC-QCL system also probed for polarization-dependent losses in the single-crystal optical coatings.

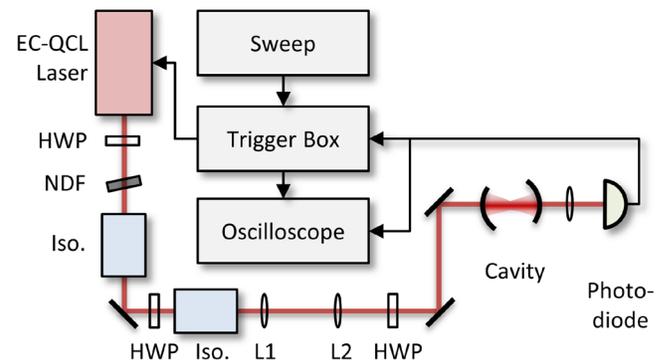

**Fig. 3.** Schematic of the ring-down setup at NIST. EC-QCL, external-cavity quantum cascade laser; HWP, half-wave plate; NDF, neutral density filter; Iso., optical isolator.

To construct the cavity at NIST, each mirror was secured at the center of a vacuum viewport using a 25.4 mm diameter retaining ring and isolated from the laboratory environment using wedged CaF$_2$ vacuum windows. The mirrors were tested under vacuum while experiencing zero pressure difference between their AR- and HR-coated faces. Nominal mirror separation defining the optical cavity length was estimated from machine drawings to be $d = 15.6 \pm 0.6$ cm and confirmed by caliper measurements of the individual components prior to assembly. The cavity free spectral range was therefore $\nu_{FSR} = 960 \pm 40$ MHz.

Frequency-dependent total losses were measured by coarse temperature tuning of the continuous-wave output of the EC-QCL. At each unique frequency $\nu$, cavity transmission during laser current dithering was monitored by a liquid-nitrogen-cooled InSb photodetector. Once transmission reached a user-defined threshold, the laser current was further shifted by summing a square wave signal with the dither signal to rapidly extinguish the pumping laser field, thus yielding a single decay event which was amplified and fitted for the exponential decay time constant $\tau$. A block diagram of the experimental set-up is shown in Fig. 3.

Optical components shown in Fig. 3 were placed at etalon immune distances [35] (whenever possible) and tilted relative to the incident laser beam to effectively eliminate the coupling of scattered light with the optical cavity mode. To avoid the coupling of spurious reflections from the AR-coated face of the mirrors into the optical cavity mode, the two-mirror cavity was aligned slightly off-axis, creating an effectively wedged surface at the flat AR-coated face relative to the HR coated concave mirror surface. Detailed methods of off-axis cavity alignment are reported in the supplementary materials.

**Direct Transmission Measurements**

To experimentally decompose the individual mirror total loss components summarized in $1-R$, we used a ratiometric, lock-in amplified method to directly probe mirror transmittance. The underlying principle of this measurement is to compare the incident and transmitted intensities for a single mirror using the same FP-QCL source as in the CDL ring-down measurements. Key elements of this setup are the high sample irradiance offered by the laser source, lock-in detection to increase the SNR and dynamic range of the detector, as well as ratiometric detection to account for power fluctuations of the laser during acquisition. Given the high output power, spectrally resolved measurements could be achieved by using the grating monochromator.

The experimental apparatus is based on a FP-QCL in quasi-continuous wave (QCW) mode (refer to Fig. 4 for abbreviations), controlled by the diode controller's (Thorlabs ITC4002QCL) internal QCW modulation. The QCW square-wave signal was used as a reference for the digital multi-channel lock-in amplifier (LIA, model Zurich Instruments HF2LI). The monochromator was used to narrow the measurement range to a passband of < 2.5 nm for each measurement. This output was then divided into two separate beam paths by a plate BS to allow monitoring of intensity fluctuations at the measurement wavelength. A half-wave plate and a polarizer (ISP Optics POL-1.5-5-SI) in front of the BS were used to set p-polarization with respect to the BS, thereby avoiding any changes in its splitting ratio. In the sample path, leading to D1, we used a 2× beam reducer to obtain a smaller beam diameter when probing the HR sample.

The two identical detectors (Thorlabs PDA20H) were connected to separate channels of the LIA and simultaneously recorded during the measurement. By dividing the respective demodulated amplitudes CH1/CH2 we suppressed any common fluctuations of the input, obtaining a time trace with drastically reduced variations. Following this approach at each measurement wavelength, a pair of measurements with and without the HR sample was taken. By dividing the mean value of each, we obtained the intensity transmission coefficient in the monochromator bandpass.

Care was taken to avoid any perturbation of the monitor path or the FP-QCL by placing the HR sample at a slight angle to avoid systematic errors in $T$ due to back-reflections. Two ground glass diffusers were placed in front of the detectors to suppress the influence of spatial variation of responsivity across the active area of the detector [36]. To ensure linear detector response over the measurement range of five orders of magnitude at the targeted uncertainty levels, the QCL output was attenuated via neutral density filter (NDFs) while the LIA served to sufficiently increase detector sensitivity and dynamic range.

Accuracy in these measurements was largely limited by beam steering effects related to the insertion of the mirror sample and estimated to be ±12 ppm. Comparative relative measurements (e.g. as a function of input polarization) could however benefit from the significantly better precision of ±2 ppm.

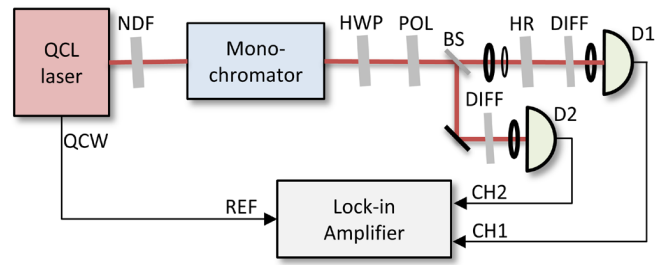

**Fig. 4.** Block diagram of the direct transmission setup. QCW, quasi-continuous wave modulation signal; NDF, reflective neutral density filter; HWP, half-wave plate; POL, polarizer; BS, 50/50-coated beamsplitter; HR, sample under test; DIFF, ground glass diffusor; D1/D2, detectors.

**Direct Absorption Measurements**

For independent absorption measurements, the FP-QCL was again used as a pump laser (capable of 450 mW output power) in a PCI system (Stanford Photo-Thermal Solutions) modified for mid-IR operation. In this setup, absorption in the thin film coating at the pump wavelength was measured indirectly by probing the thermal lens induced in the non-absorbing substrate via a second probe laser [37]. Both pump and probe lasers were focused onto the same spot of the sample coating (with waist diameters of about 70 μm and 200 μm, respectively). The localized "hot spot" induced by the more tightly focused pump imprints a phase distortion on the central part of the more loosely focused probe beam, causing interference with the undisturbed outer probe wave front (which acts as the reference arm of the common-path interferometer). The resulting interference pattern appears in the near field of the interaction region, the central maximum of which is then imaged onto a detector. Using lock-in detection referenced to the chopped pump laser, the relative intensity modulation depth $\Delta I/I$ of the probe signal was recorded. Normalized by the incident pump power $P$, the signal is proportional to absorption $\alpha$ over several orders of magnitude, with proportionality constant $R_c$ denoting the system responsivity: $\Delta I/(IP) = \alpha R_c$. PCI absorption measurements are a common characterization technique in the near-IR. Extending this technique to the mid-IR however involved overcoming numerous challenges related to these longer wavelengths, including the substrate response, probe wavelength, detection path of the probe (i.e. in transmission or reflection), and calibration technique.

In the near-IR, $R_c$ is determined by measuring a calibration sample of known absorption. This calibration step requires that the substrate material as well as the absorbing surface layer thickness match the sample under study to assure a comparable thermal response. Typically, a metallic coating with broadband absorption of a few tens of percent serves as such a reference sample. However, as a consequence of mid-IR thin film coatings being rather thick (owing to the long wavelength and corresponding increase in the individual quarter-wave layer thicknesses), the thermal lens in the coating layer itself becomes non-negligible, making it challenging to fabricate such a calibration sample with comparable thermal response. It is for this reason that we resorted to a novel in-situ calibration technique employing a more strongly absorbed "proxy pump" laser [37] with a wavelength of 532 nm (above the GaAs bandgap) in combination with the sample itself, to provide the reference absorption measurement. In either case, the high reference absorption can be independently determined via direct measurements of incident, transmitted, and reflected power, with the relative error of the calibration measurement transferring to the measurement of the actual sample of much lower (typically ppm-level) absorption. Care was taken to focus the proxy pump to the same diameter as the primary pump, to ensure the same geometry of the thermal lens. This was confirmed by assuring that the same value of $R_c$ is deduced for both pumps when using a custom-made calibration sample of similar known high absorbance for both pump wavelengths (broadband 7 nm Cr overcoat on a $CaF_2$ substrate).

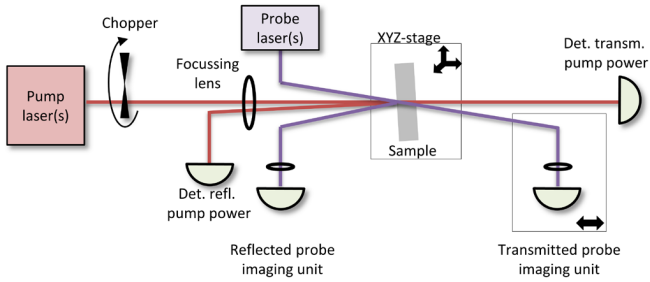

**Fig. 5.** Block diagram of the photothermal interferometer for absorption measurements (PCI). A slight angle between pump and probe beams allows dumping of the pump and pickup of either transmitted or reflected probe beam.

Using fused silica (FS) as a substrate material typically offers a high system responsivity $R_c$ and therefore high sensitivity, owing to its low thermal conductivity and the strong temperature dependence of the refractive index. Although a dedicated sample with the 4.54 µm crystalline coating bonded to FS was available, it was found that the heating contribution from substantial absorption of the transmitted mid-IR pump light within the FS substrate itself was not sufficiently distinguishable from pump absorption within the actual coating. For this reason, we resorted to the use of Si substrates (transparent at the pump wavelength) in all further measurements, at the expense of lower SNR (given the high thermal conductivity of the substrate). Keeping in mind that PCI is an interferometric method, the use of short wavelength probe light is favored for high system responsivity. Measurements were therefore conducted with a HeNe probe at 633 nm in reflection. However, with a probe photon energy (1.96 eV) exceeding the GaAs bandgap energy of 1.42 eV, the measured pump absorption is distorted by additional probe-induced absorption via free carriers. We therefore took several measurements at decreasing probe power and extrapolating to zero probe power [15].

The crystalline mirror coatings were repeatedly purged with dry nitrogen between measurements to mitigate influences of surface contamination. To perform measurements at the required ppm-level precision, it was necessary to ensure a heat-up time of the QCL pump laser of around 15 minutes to maintain a constant power level and beam profile.

## 3. RESULTS

**Total loss and transmittance**

Our measurement results, summarized in Fig. 6 and Table 1, show excellent performance for these prototype mirrors at 4.54 µm. The refined transmission model derived from XRD, cross-sectional SEM, and FTS is shown as the solid red curve in Fig. 6 (top). The above-described routine allows for an estimate of the center wavelength $\lambda_{0,T} = 4\,538 \pm 1$ nm and an expected mirror transmittance $T(\lambda_{0,T}) = 144 \pm 2$ ppm at the stopband center, with the error bands obtained by the same procedure. The shift of the measured center wavelength from the design target of 4 500 nm is due to unavoidable deviation in layer thicknesses by a small global scaling factor during MBE growth (< 1% for the studied samples). It can be readily seen from Fig. 6 (top) that our model, based on the as-grown layer geometry derived from SEM imaging (capturing thickness variations of different high and low index layers with a relative standard deviation of 2%), captures modulations in the side lobes observed in FTS much better than a model assuming a strictly periodic layer structure (dashed red design curve). Although we assumed an error of 1% for XRD measurements of the Al alloy composition, our evaluation suggests a much lower uncertainty of 0.2%, with best-fit values for the Al composition ranging only from 0.9229 to 0.9247. Comparing these results with our direct transmission measurements (Brown triangles in Fig. 6 (mid) and Fig. 6 (bottom)), we observe excellent agreement within the error bounds (reflecting the full uncertainty budget).

**Table 1.** Comparison of design and experimental values for cavity end-mirrors with mid-infrared crystalline coatings. For total loss 1−R and central wavelength $\lambda_0$, the mean value of different measurements of Fig. 6 is presented. As mentioned in the introduction, optical scatter is negligible in this case and thus its absolute value and uncertainty is ignored. Design values for absorption A are taken from Fig. 7 in Ref. [15]. Actual sample absorption is inferred from all other losses.

| Parameter | Design | Measured | Uncertainty | Units |
|---|---|---|---|---|
| $\lambda_0$ | 4 500 | 4 535 | 1 | nm |
| 1-R | <162 | 151 | 3 | ppm |
| T | 142 | 144 | 2 | ppm |
| S | <1 | -- | -- | ppm |
| A | <20 | 7 | 4 | ppm |

Total loss values (measured with the incident linear polarization state set to $\phi = 90°$) extracted from the ring-down measurements at CDL (dark blue curve, minimum of $153 \pm 1$ ppm at $\lambda_{0,CDL} = 4\,534 \pm 1$ nm) and NIST (light blue data points, minimum of $149 \pm 6$ ppm at $\lambda_{0,NIST} = 4\,533 \pm 1$ nm) are also included. For both, the relative standard uncertainty in the total optical losses, $u_r = \sigma_{1-R}/(1-R)_{min}$, was estimated from a quadrature sum of type-A reproducibility and a conservative estimate of dominating type-B systematic uncertainty (mainly dictated by the cavity length uncertainty). The standard uncertainty at the center wavelength of ±1 nm was taken to equal the accuracy of monochromator calibration (CDL) and accuracy specified by the laser manufacturer (NIST).

From the average value of $(1-R)_{min}$, presented in Table 1, we infer a cavity finesse of $\mathcal{F} = 20\,805 \pm 413$. This is a 5× improvement over the first mid-infrared monocrystalline mirror coatings with $\lambda_0 = 3.7$ µm [15], and the maximum cavity transmission is now improved to $T_{cav} > 92\%$ from a value of 24% that would result from the assumption of $L = 160$ ppm, based on [15]. Thus, we achieve a modest cavity finesse, in agreement with the multilayer stack design chosen for these first prototype mirrors, while the throughput is significantly enhanced by the extremely low level of excess losses.

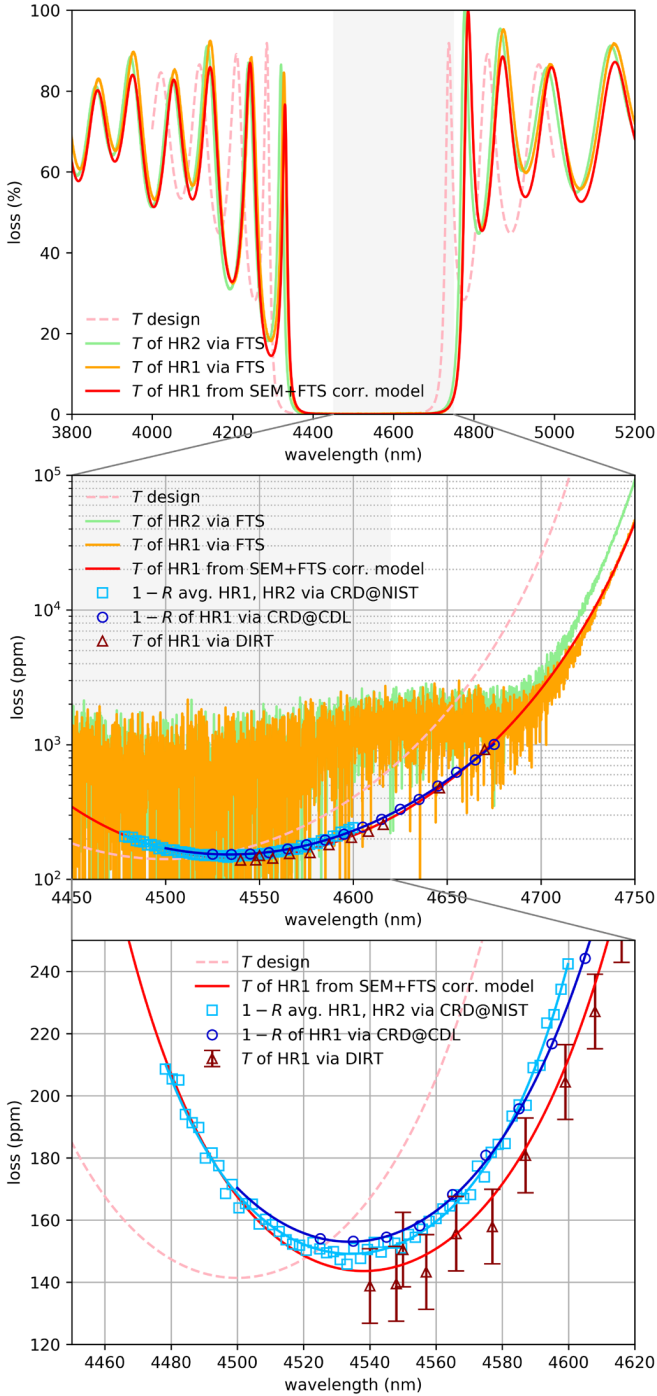

panel: Zoom in to the stopband center showing the low SNR of the FTS results. The derived model (red line) shows good agreement with all other high-precision loss measurements. Bottom panel: A detailed look at the total loss and transmittance. The total loss (plotted in shades of blue) was independently measured by two separate teams using different realizations of a cavity ring-down scheme (CRD) and combinations of mirrors with nominally identical minimum transmittance from the same production batch. Data sets are fitted with 4th-order polynomials to guide the eye (solid lines). The transmittance was determined from the model calculation (red line) and verified by direct transmission measurements (DIRT, brown triangles).

**Mirror birefringence**

The NIST setup was used to measure mirror birefringence by performing frequency-swept CRD at an incident linear polarization of $\phi = 45°$, exciting both orthogonal birefringence modes during the frequency sweep [38]. To observe beating from the orthogonal birefringence modes, a linear polarization analyzer was placed after the cavity and before the photodetector at a matching relative angle of $\phi = 45°$. The resulting decays deviated strongly from exponential behavior, as evidenced by the representative cavity decay plotted in Fig. 7. Birefringence mode beating is clearly a large perturbation on the single-mode decay, and its frequency is resolved by our detection system with > 10 MHz of bandwidth. Using the most general mode beating equations discussed in Ref. [38], we fitted the cavity decay event at $\phi = 45°$ to yield a birefringence splitting of $\Delta \nu_{beat} = 110 \pm 5$ kHz, a value larger than the inferred cavity mode linewidth of $\delta \nu_{cav} = 45.6 \pm 1.8$ kHz. The observed birefringence is presumably induced by anisotropic strain in the crystalline multilayer as discussed below.

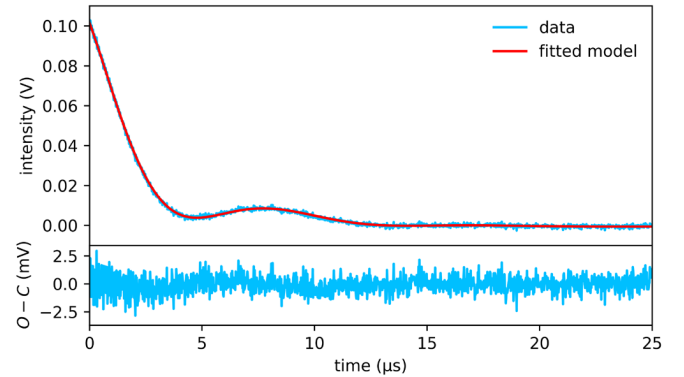

**Fig. 7.** Typical cavity ring-down (CRD) trace with visible polarization mode beating, recorded at NIST for an input polarization angle of $\phi = 45°$. An associated birefringence splitting of $\Delta \nu_{beat} = 110 \pm 5$ kHz is extracted from the fitted model (corresponding residuals shown at the bottom).

**Observation of polarization-dependent absorption loss**

While sources of systematic error in the current mid-IR PCI setup prohibit absolute absorption measurements at the desired few-ppm level of sensitivity, we were still able to observe a very distinct and unexpected relative change of absorption loss as a function of incident pump laser polarization. As shown in Fig. 8, absorption reaches a minimum for linear pump polarization oriented at a relative angle of $\phi = 90°$ with respect to the $[0\bar{1}\bar{1}]$ crystal axis (represented by the coating flat) and a maximum for $\phi = 0$. It was verified that such an effect is only observed for crystalline coatings (by comparison to standard PVD-coated mid-IR high reflectors). Analysis of high-finesse 1 550 nm

**Fig. 6.** Summary of spectrally resolved results for total loss $1−R$ and transmittance $T$. All measurements were performed for incident linear polarization angle $\phi = 90°$ relative to the $[0\bar{1}\bar{1}]$ crystal axis. Top panel: The two sample mirrors are distinguished according to their broadband transmission spectra (labelled HR1 and HR2) and differ by a slight shift of center frequency (likely due to inhomogeneities during layer growth). The solid red line represents a model for the as-grown layer structure of HR1, based on a fit to the FTS results, and layer thicknesses derived from SEM images, accurately reproducing the asymmetric structure of side-maxima outside the stopband. The thin dashed line shows the model calculation for uniform target layer thicknesses for comparison. Middle

crystalline coatings appear to show similar effects, however given the low absorption in those coatings (at the ~ 1 ppm level) the change of absorption is difficult to accurately quantify (T. Legero, PTB, Germany, personal communication, February 2019). We excluded that the effect is an artifact of the measurement geometry by verifying that rotation of the sample produced the same effect as rotating the pump polarization. It was also observed that the effect has no appreciable dependence on small changes to the pump angle of incidence.

As evidenced by the blue data points in Fig. 8, the same polarization dependence was observed in CRD measurements at NIST, where relative peak-to-peak variation in total loss was quantified to be 8.3 ppm. No dedicated polarization analyzer was used between the cavity and the detector to avoid the previously discussed effect of birefringence mode beating. A complete second wavelength-dependent set of $1-R$ values was also calculated from CRD measurements with the incident linear polarization state $\phi = 0°$. Compared to the orthogonal configuration plotted in Fig. 6, the $\phi = 0°$ data set yielded a slightly higher value of $(1-R)_{min} = 158 \pm 6$ ppm at a nearly identical center wavelength of $\lambda_0 = 4\,532$ nm, in good agreement with the polarization-dependent changes in total losses observed in Fig. 8.

No such polarization dependence was observed in direct transmission measurements (brown triangles in Fig. 8). Since scatter is expected to be bounded to the single ppm level (as determined for comparable near-IR crystalline mirrors), and direct absorption measurements are insensitive to scatter, all observations suggest that the observed polarization dependent loss can be solely attributed to absorption.

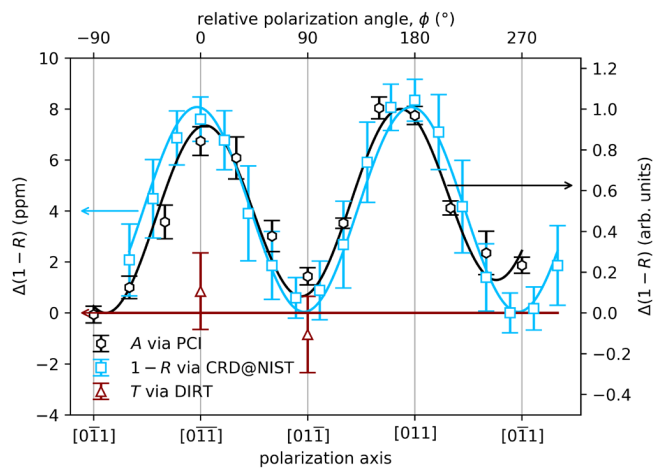

**Fig. 8.** Change in various optical losses plotted vs. linear polarization angle $\phi$ relative to the $[0\bar{1}1]$ crystallographic plane. About 8 ppm peak-to-peak variation is measured in total loss (1-R) cavity ring-down (CRD) measurements between orthogonal input polarization states, reproducing the qualitative behavior observed for absorption $A$ via photothermal common-path interferometry (PCI). Direct transmission measurements (DIRT) suggested no polarization dependence in coating transmittance.

**Theoretical modelling of polarization-dependent absorption and refractive index**

From our current understanding of these novel low-loss monocrystalline mirrors, free-carrier absorption is the limiting mechanism for absorption loss for below-bandgap illumination. This is supported by theoretical estimates of the limiting loss for acceptor-dominated materials with relevant background impurity levels. Initially, it was assumed that lattice-mismatch-mediated strain was the driving force behind both the birefringence (Re{$\Delta n$}) and the orientation-dependent absorption (governed by Im{$\Delta n$}). However, owing to the cubic nature of the zincblende unit cell, an in-plane biaxial strain for (100)-oriented GaAs, as implemented here, will lower the symmetry from cubic to tetragonal. The optical axis will be along the illumination direction, so that no birefringence or absorption differences are possible. Nevertheless, birefringence occurs in our crystalline mirrors at a variety of wavelengths with apparently repeatable magnitude. It is assumed that a similar underlying process drives the anisotropy in both the index and free-carrier absorption.

A potential candidate for causing the observed polarization-dependent absorption is anisotropic strain relaxation as observed in InGaP-based materials [39] which breaks the system symmetry. To test this hypothesis, we simulated the free-carrier absorption caused by an additional 1% strain along the $[0\bar{1}1]$ direction. While we do not expect that such large additional strains are present in this system, the large imparted strain value facilitates the convergence of the simulations.

For these simulations we used density functional theory with the PBEsol exchange-correlation functional [40], as implemented in the Quantum-ESPRESSO package [41]. The electron-phonon matrix elements were modeled by a Fröhlich model, which has been shown to work well within the infrared region [42–44]. We used a rotated conventional unit cell, an energy cutoff of 100 Ry, and a 14×20×20 k-point grid. Spin-orbit coupling was not included. The refractive index was determined using a 50×50×50 Wannier interpolated [45] k-point grid as implemented in the WANNIER90 package [46]. The underlying electronic structure was calculated using the HSE06 [47] hybrid functional, as implemented in the VASP package [48].

Fig. 9 shows the calculated absorption cross section (the absorption coefficient divided by the free-carrier concentration) as a function of the illumination angle for photons with an energy of 1.2 eV (corresponding to a wavelength of approximately 1.0 μm). As with the large uniaxial strain value, a higher energy is probed to simplify these proof of principle calculations. Note that similar behavior is observed for other photon energies. The calculations show a polarization dependence similar to Fig. 8, indicating that anisotropic strain can cause the observed effect.

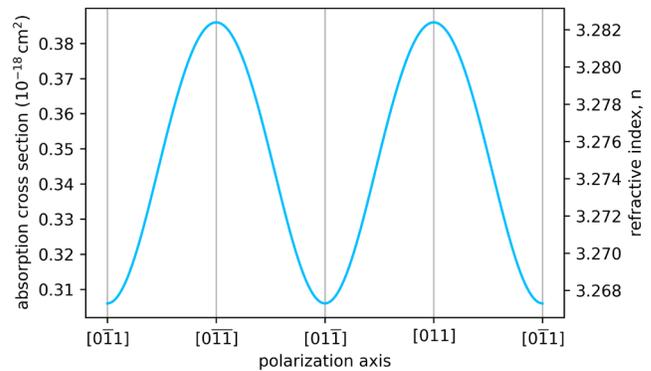

**Fig. 9.** Simulation of the orientation-dependence of the photon absorption cross section and refractive index of uniaxially strained GaAs as a function of linear polarization direction. Results are given for a photon energy of 1.2 eV and in-plane compressive strain of 1% along the $[0\bar{1}1]$ direction. These values are chosen to expedite these proof of principle calculations.

## 4. Conclusions and Outlook

We successfully fabricated and optically characterized state-of-the-art crystalline interference coatings at 4.54 µm. Optical losses were determined using a variety of experimental approaches to ultimately decompose coating transmittance, absorption, and scatter. Two independent cavity ring-down systems were used to determine the total loss, while an extension of the PCI measurement scheme to the mid-IR using a novel in-situ "proxy pump" calibration technique was used to determine polarization-dependent relative changes in absorption loss. Transmittance was measured directly and calculated using a transmission matrix model based on measurements of as-grown layer thicknesses and broadband FTIR spectra.

The fact that many of the outlined measurement methods are based on a single FP-QCL makes the measurements easily extendable to a wide variety of wavelength regions. This lays the groundwork for their routine application in a production environment and will speed up extension of crystalline coating technology into the molecular fingerprint region.

In the course of our efforts, we observed a polarization dependence of absorption losses in the studied crystalline multilayer stacks, which, to the best of our knowledge, has not yet been reported in the literature. Initial proof of principle calculations show that anisotropic strain can lead to polarization-dependent free-carrier absorption and a polarization dependent refractive index. More detailed experimental and theoretical investigations of this effect are in progress to determine the root cause of this anisotropic strain relaxation.

Our measurements confirm record-low levels of excess optical losses (scatter plus absorption) below 10 ppm, with $A + S = 7 \pm 4$ ppm in these coatings. With such low levels of excess loss, we have now demonstrated mid-IR mirrors capable of optical performance on par with their near-IR counterparts. This offers a bright outlook for diverse applications in the mid-IR spectral region, e.g. in future cavity-enhanced spectroscopy applications covering the molecular fingerprint region, laser stabilization, fundamental physics experiments as well as many other applications. In follow-on efforts it should be possible to produce optical resonators with center wavelengths in the range of 2 to 5 µm and a cavity finesse exceeding 100 000, a significant milestone in the development of the first mid-IR "super mirrors". Such exceptional levels of performance can be achieved via a minor design alteration, specifically reducing the transmittance of the interference coating to below 10 ppm. TMM calculations show that a transmittance of 8 ppm can be achieved at 4.5 µm with a full mirror stack of ∼ 33 µm in thickness (45.5 periods of GaAs/Al$_{0.92}$Ga$_{0.08}$As). Using the stacking approach employed here, the thickness of individual half mirror stacks would be below 17 µm, with state-of-the-art MBE generating a surface quality sufficient for direct bonding. We also note that this process would yield a coating surface quality (post substrate transfer) with an RMS surface roughness below 0.2 nm, rendering optical scatter negligible. The absorption should be unaffected also, as there would be no change in the level of background doping, nor in the optical penetration depth into the multilayer.

**Funding**. Austrian Federal Ministry for Digital and Economic Affairs; National Foundation for Research, Technology and Development; Christian Doppler Research Association; Austrian Research Promotion Agency (865556); National Institute of Standards and Technology; National Research Council Postdoctoral Research Associateship Program; Austrian Science Fund (M2561-N36).

**Acknowledgments.** We acknowledge support by the Faculty Center for Nano Structure Research at the University of Vienna in supplying the infrastructure for SEM imaging. A portion of this work was performed in the UCSB Nanofabrication Facility.

**Disclosures.** GW, LWP, ASM, JF, OHH, GZ, DMB, AJF: Thorlabs Inc. (F); GDC, GWT, DB, DF, PH, CD: Thorlabs Inc. (E). Certain commercial equipment is identified in this paper in order to specify the experimental procedure adequately. Such identification is not intended to imply recommendation or endorsement by the National Institute of Standards and Technology, nor is it intended to imply that the equipment identified is necessarily the best available for the purpose.

See Supplement 1 for supporting content.

# Mid-infrared monocrystalline interference coatings with excess optical loss below 10 ppm: supplemental document

**Prevention of etaloning in swept-frequency cavity ring-down measurements**

The external-cavity quantum cascade laser (EC-QCL) swept-frequency cavity ring-down instrument at the National Institute of Standards and Technology achieved single cavity-mode excitation, and therefore was capable of resolving stray etalons formed by spurious reflections involving one of the cavity mirrors. To reduce the influence of these coupled cavity effects [49], broader frequency scans of the EC-QCL were performed by increasing the intensity of the laser current dither to cover multiple longitudinal modes. Cavity decay signals were then triggered at each longitudinal mode and values of $\tau$ were recorded, searching for periodic modulation in the apparent optical losses which would indicate a coupled cavity. With the optical cavity aligned on-axis, the measured values of $\tau$ revealed one strong etalon with a characteristic frequency of $\nu_{\text{etalon}} \approx 6.4\nu_{\text{fsr}} = 6.1$ GHz, as depicted in Fig. S1 (left). This value was in reasonable agreement with an etalon formed by the AR coated faces of the mirrors themselves, located on the opposite sides of their silicon substrate with $d = 0.635$ cm and material index $n \approx 3.42$. (i.e. 6.9 GHz).

The absence of significant etaloning with those optical components was confirmed by the inclusion of additional neutral density filters before and after the cavity to reduce the coupled cavity finesse, a procedure which had no effect on the etalon shown in Fig. 3. We then performed the following additional cavity alignment steps which effectively eliminated the etalon. First, the photodetector position was translated in the plane of the optical table using a micrometer and linear optomechanical stage. With the new photodetector location acting as a fixed target, laser alignment to the cavity was adjusted until high throughput and good mode matching conditions were again achieved. The procedure forced the cavity alignment to slowly walk off-axis, or off center of the planar-concave mirror substrates, effectively creating a wedged substrate by reducing parallelism between the planar AR- and concave HR-faces as depicted in Fig. S1 (right). The procedure was repeated until the longitudinal mode frequency scan showed only minor variations in the relative fractional time constant $\epsilon_\tau = \tau/\bar{\tau} - 1$, where $\bar{\tau}$ is the average ringdown time over all longitudinal modes.

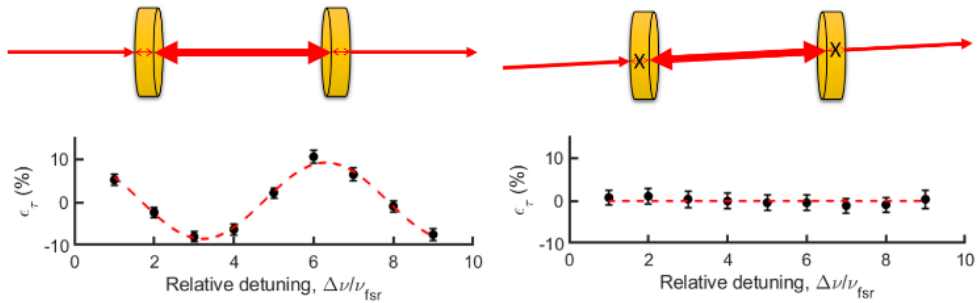

Fig. S1. Cavity alignment for the reduction of parasitic etalons. Left: Illustration of on-axis cavity alignment and observed etalon attributed to coupling of scattered light between the AR and HR mirror faces. Right: Equivalent depiction of off-axis cavity alignment and thereby reduced AR-HR face etalons. In both cases, $\epsilon_\tau = \tau/\bar{\tau} - 1$ is the fractional change in $\tau$ relative to the average value $\bar{\tau}$.